\documentclass[onecollarge]{svjour2}       
\usepackage{mathptmx}      
\usepackage{amsmath}
\usepackage{amssymb}
\usepackage{cite}
\journalname{Foundations of Physics}
\begin{document}

\title{On the Geodesic Nature of Wegner's Flow
\thanks{S.A. was supported in part by a Grant-in-Aid for Scientific Research from the Japan Society for the Promotion of Science.}
}


\author{Yuichi Itto         \and
        Sumiyoshi Abe 
}


\institute{Y. Itto \at
              Science Division, Center for General Education, 
              Aichi Institute of Technology,
              Aichi 470-0392, Japan
 \\
              \email{itto@aitech.ac.jp}           
           \and
           S. Abe \at
              Department of Physical Engineering, Mie University, Mie 514-8507, Japan\\
              \email{suabe@sf6.so-net.ne.jp}
}

\date{Received: date / Accepted: date}

\maketitle

\begin{abstract}
Wegner's method of flow equations offers a useful tool for 
diagonalizing a given Hamiltonian and is widely used in 
various branches of quantum physics. Here, generalizing 
this method, a condition is derived, under which 
the corresponding flow of a quantum state becomes geodesic 
in a submanifold of the projective Hilbert space, 
independently of specific initial conditions. 
This implies the geometric optimality of the present method 
as an algorithm of generating stationary states. 
The result is illustrated by analyzing some physical 
examples.
\end{abstract}

\section{Introduction}
\label{intro}
\ \ \ \ \ In recent years, much attention has been focused on 
Wegner's method of flow equations \cite{Wegner1} 
(for reviews, see \cite{Wegner2}). 
This method offers a powerful tool for diagonalizing a given 
quantum-mechanical Hamiltonian, which has been devised in 
analogy with the theory of renormalization groups and has 
widely been applied to a variety of problems in condensed 
matter physics, nuclear and particle physics, and 
quantum information. Examples include the effective 
Hamiltonian of the Anderson impurity model \cite{Kehrein}, 
a Dirac particle in an external electromagnetic 
field \cite{Bylev}, 
an effective spin-spin coupling arising from spin-phonon 
chains \cite{Rass}, the $t-t'$ Hubbard model for 
high-temperature superconductivity \cite{Hankevych}, 
the Tomonaga-Luttinger model for interacting spinless 
electrons in one dimension \cite{Stauber}, 
localized superfluidity \cite{Domanski}, 
the Lipkin-Meshkov-Glick model in nuclear physics \cite{Dusuel1},
 quantum phase transition in the interacting boson 
model \cite{Dusuel2}, light-cone quantum 
chromodynamics \cite{Glazek}, electron-electron and 
electron-phonon interactions in the Hubbard-Holstein 
model \cite{Aprea}, and stationary photon-atom 
entanglement \cite{Itto}.

Wegner's method employs a continuous unitary 
transformation represented by the operator, $U(l)$, 
where $l \in [0,\infty)$ is referred to as the flow parameter.
$U(l)$ transforms the original Hamiltonian $H=H(0)$ 
to $H(l)=U(l)HU^{\dagger}(l)$, which satisfies 
$dH(l)/dl=[\eta(l),\ H(l)]$, where $\eta(l)$ is 
the anti-Hermitian generator given 
by $\eta(l)=\bigl[dU(l)/dl\bigr]U^{\dagger}(l)$. 
Wegner's choice for $\eta(l)$ to diagonalize 
(or, block-diagonalize) the Hamiltonian reads
\begin{equation}
\eta(l)=\eta^{W}(l) 
\equiv \bigl[ H_{\text{d}}(l),\ H_{\text{o-d}}(l) \bigr],
\end{equation}
where $H_{\text{d}}(l)$ and $H_{\text{o-d}}(l)$ stand for 
the diagonal and off-diagonal parts of $H(l)$, respectively. 
It can be shown \cite{Wegner1,Wegner2} 
that $H_{\text{o-d}}(l)$ tends to vanish 
in the limit $l \rightarrow \infty$. 
Mathematically, this procedure corresponds to 
the Jacobi algorithm for eigenvalue problems, 
which defines a steepest-descent flow in the space of 
matrices with the Frobenius norm \cite{Chu}.
	
Although Wegner's method has been applied to many problems 
as mentioned above, its inherent properties do not seem to 
be explored well. It is our opinion that this method is more 
than just a mathematical tool for diagonalizing a Hamiltonian. 
The purpose of this paper is to reveal an interesting geometric
 property hidden behind Wegner's flow equations. For it, 
we first generalize Wegner's method itself. 
The original method of Wegner is a special case within 
this new framework. Next, we consider the flow of a given 
quantum state through a submanifold of the projective Hilbert 
space composed of rays explained below. 
Then, we find a condition, under which the flow becomes 
geodesic in the submanifold, 
\textit{independently of specific ``initial'' conditions at} 
$l=0$. This remarkable property is illustrated by 
analyzing some physical examples. 
Our discussion may also have significance for 
laboratory experiments, since one often need realize 
a stationary state from an initially prepared state 
via quantum operations \cite{Itto}, 
and in this context the unitary transformation in 
Wegner's method is now shown to offer the optimal strategy 
for it in a certain class of systems 
(satisfying the condition discovered here).

\section{Generalization of Wegner's Method of Flow Equations}
\label{sec:Generalization}

\ \ \ \ \ Let us start our discussion with considering a quantum system 
in a $d$-dimensional Hilbert space, where $d$ is either 
finite or infinite. 
Its unitary-transformed Hamiltonian, $H(l)$, is decomposed 
into two parts: $H(l)=H_{\text{d}}(l)+H_{\text{o-d}}(l)$, 
where $H_{\text{d}}(l)$ and $H_{\text{o-d}}(l)
=\sum_{a=1}^A \ H^{(a)}_{\text{o-d}}(l) 
\quad (1 \leq A \leq d-1)$ 
are the diagonal and off-diagonal parts, respectively. 
Using the complete set of the normalized eigenstates 
of $H_{\text{d}}(l)$, $\{ \mid u_{n} \rangle \}_{n=1,2,\dots,d}$, 
which is a \textit{natural basis} such as the Fock basis of 
the harmonic oscillator, 
we write them in the following forms:
\begin{equation} \label{diagonal}
H_{\text{d}}(l)=\sum_{n=1}^d \epsilon_{n}(l) 
\mid u_{n} \rangle \langle u_{n} \mid,
\end{equation}
\begin{equation} \label{ath off-diagonal}
H^{(a)}_{\text{o-d}}(l)
=\sum_{n=1}^d 
\Bigl[C^{(a)}_{n+i_{a}}(l) \mid u_{n+i_{a}} \rangle \langle u_{n} \mid
+C^{(a)*}_{n+i_{a}}(l) \mid u_{n} \rangle \langle u_{n+i_{a}} \mid \Bigr],
\end{equation}
where $\epsilon_{n}(l)$ is the $n$th eigenvalue 
of $H_{\text{d}}(l)$ and $C^{(a)}_{n+i_{a}}(l)$'s are 
the complex expansion coefficients. 
The nonzero index $i_{a}$ describes the off-diagonality 
and is ordered without loss of generality 
as follows: $0<i_{1}<i_{2}< \cdots <i_{A}$. 
It is understood that $C^{(a)}_{n+i_{a}}(l)=0$
($C^{(a)}_{n-i_{a}}(l)=0$), if $n+i_{a}>d$  ($n-i_{a} \leq 1$ ).
	
Here, we generalize Wegner's method as follows. 
Instead of taking the whole of $H_{\text{o-d}}(l)$, 
we employ only a single term, say $H^{(a)}_{\text{o-d}}(l)$:
\begin{equation} \label{ath eta}
\eta^{(a)}(l)=
\Bigl[H_{\text{d}}(l),\ H^{(a)}_{\text{o-d}}(l)\Bigr].
\end{equation}
Clearly, Wegner's choice 
is $\eta^{W}(l)=\sum_{a=1}^A \ \eta^{(a)}(l)$. 
It can be found after a straightforward calculation 
using Eqs. (\ref{diagonal})-(\ref{ath eta}) that 
the corresponding generalized flow equation
\begin{equation} \label{generalized flow equation}
\frac{d H(l)}{d l}=\bigl[\eta^{(a)}(l),\ H(l)\bigr]
\end{equation}
gives rise to
\begin{align} \label{decay}
\frac{d}{d l}\sum_{m,n=1 \ (m \neq n)}^d
| \langle u_{m} \mid H(l) \mid u_{n} \rangle |^2
=&-\frac{d}{d l}\sum_{n=1}^d \epsilon_{n}^2(l) \nonumber \\
=&-4\sum_{n=1}^d 
\Bigl[ \epsilon_{n+i_{a}}(l)-\epsilon_{n}(l) \Bigr]^2 
\Bigl| C^{(a)}_{n+i_{a}}(l) \Bigr|^2,
\end{align}
implying that the off-diagonal elements of $H(l)$ tend to 
decay as $l$ increases. 
(The first equality in Eq. (\ref{decay}) comes from the 
fact that $(d/dl)\text{Tr}[H^2(l)]=0$.)
$\eta^{W}$ is obtained from the set 
$\{ \eta^{(a)} \}_{a=1,2,\dots,A}$, but the reverse is 
obviously not possible if $A \geq 2$. 
In this sense, the present method is regarded 
as a generalization of Wegner's.
	
Closing this section, we emphasize the following point. 
Like in Wegner's method, reduction of off-diagonal elements 
of a Hamiltonian does not necessarily mean exact vanishing 
of the elements. In Wegner's case, it tends to be 
block-diagonalization of a Hamiltonian, in general, 
whereas Eq. (\ref{decay}) implies reduction 
of \textit{targeted} off-diagonal elements, 
not all off-diagonal elements. 
The present generalized method enables one to look into 
the details of the structure of a Hamiltonian.

\section{Condition for Geodesic Flow}
\label{sec:Condition}
\ \ \ \ \ Next, let us translate the flow of the Hamiltonian into 
the flow of a state. 
Given a normalized state $\mid \psi \rangle$, 
the stationary Schr\"{o}dinger equation reads 
$H \mid \psi (\infty) \rangle=E\mid \psi (\infty) \rangle$, 
where $\mid \psi(l) \rangle=U^{\dagger}(l)\mid \psi \rangle$. 
As well known, two states different from each other only by 
total phases are equivalent in quantum mechanics, 
and therefore a physical state is represented by a ray. 
Accordingly, the quantum-state space is the projective 
Hilbert space that is generically a curved space. 
Eq. (\ref{generalized flow equation}) determines 
the flows of the physical coefficients contained in 
the Hamiltonian, which depend on the parameters appearing 
in the unitary operator. 
Let us explicitly write as follows: $U=U(\alpha)$, 
$\mid \psi(\alpha) \rangle=U^{\dagger}(\alpha)\mid \psi \rangle$, 
where $\alpha \equiv (\alpha^{1},\alpha^{2},\dots,\alpha^{k})$. 
The set of parameters, $\alpha$, defines a local coordinate 
on the submanifold of the projective Hilbert space. 
This submanifold is referred to as the quantum evolution 
submanifold \cite{Abe}, on which a quantum state flows. 
Then, the Fubini-Study metric \cite{Abe,Kobayashi,Page} induced on 
this submanifold is, up to the second-order infinitesimals, 
given by
\begin{equation}\label{FubiniStudy}
ds^2=1-|\langle \psi(\alpha) \mid \psi(\alpha+d\alpha) \rangle |^2
\equiv g_{ij}(\alpha) d\alpha^i d\alpha^j,
\end{equation}
where the metric tensor is expressed in terms of 
the anti-Hermitian operator 
$G_{i}(\alpha)=\bigl[\partial_{i}U(\alpha)\bigr]U^{\dagger}(\alpha)$ 
($\partial_{i}=\partial/\partial \alpha^{i};i=1,2, \dots,k$) 
as follows:
\begin{equation}\label{metrictensor}
g_{ij}(\alpha)=-\frac{1}{2}
 \langle \psi \mid 
    G_{i}(\alpha)G_{j}(\alpha)+G_{j}(\alpha)G_{i}(\alpha) \mid \psi \rangle
+\langle \psi \mid G_{i}(\alpha) \mid \psi \rangle 
              \langle \psi \mid G_{j}(\alpha) \mid \psi \rangle.
\end{equation}
Here and hereafter, Einstein's convention is understood 
for the repeated indices. The equation for a geodesic curve 
parametrized by the arc length, $\alpha(s)$, is given by 
$d^2 \alpha^{h}/ds^2+{\rm \Gamma}^{h}_{ij}(d\alpha^{i}/ds)(d\alpha^{j}/ds)=0$,
where ${\rm \Gamma}^{h}_{ij}$ is Christoffel's symbol defined by 
${\rm \Gamma}_{hij}=g_{hl}{\rm \Gamma}^{l}_{ij}=$
$(1/2)
\left(\partial_{i}g_{hj}+\partial_{j}g_{ih}-\partial_{h}g_{ij}\right)$.
	
Let us parametrize the curve by the flow parameter, 
instead of the arc length, i.e., $\alpha(l)$, and consider 
the functional
\begin{equation} \label{functional}
S[\alpha]=\int_{l_{1}}^{l_{2}} dl \ L(\alpha,\dot{\alpha}),
\end{equation}
where
\begin{equation}\label{lagrangian}
L(\alpha,\dot{\alpha})=\frac{ds}{dl}
=\sqrt{g_{ij}(\alpha) \dot{\alpha}^{i} \dot{\alpha}^{j}}
\end{equation}
and $\dot{\alpha}^{i} \equiv d\alpha^{i}(l)/dl$. 
The quantity in Eq. (\ref{functional}) is the arc length in 
the interval $[l_{1},l_{2}]$. The variation of $S$ with 
respect to  $\alpha(l)$ is calculated to be
\begin{equation}\label{variation of S}
\frac{\delta S[\alpha]}{\delta \alpha^{i}}
=-\frac{1}{4 L^3 } X_{i},
\end{equation}
where
\begin{align}\label{X}
X_{i}=&\left( 2\langle \psi \mid  \eta \mid \psi \rangle
\langle \psi \mid
\frac{\partial \eta}{\partial \dot{\alpha}^{i}} \mid \psi \rangle 
- \langle \psi \mid 
\frac{\partial \eta^2 }{\partial \dot{\alpha}^{i}}  \mid \psi \rangle 
\right)
\frac{d}{dl}\Bigl( \langle \psi \mid  \eta^2 \mid \psi \rangle
-\langle \psi \mid  \eta \mid \psi \rangle^2 \Bigr) \nonumber \\
&+2\Bigg( \langle \psi \mid 
 \left\{ \frac{d\eta}{dl},\frac{\partial \eta}{\partial \dot{\alpha}^{i}}\right\}
+\left[\eta^2, \frac{\partial \eta}{\partial \dot{\alpha}^{i}}\right]
\mid \psi \rangle
-2 \langle \psi \mid  \frac{d\eta}{dl} \mid \psi \rangle
    \langle \psi \mid  \frac{\partial \eta}{\partial \dot{\alpha}^{i}}
      \mid \psi \rangle \nonumber \\
&\hspace{1cm} -2 \langle \psi \mid  \eta \mid \psi \rangle 
  \langle \psi \mid 
     \left[\eta, \frac{\partial \eta}{\partial \dot{\alpha}^{i}}\right]
        \mid \psi \rangle \Bigg)
\Bigl(  \langle \psi \mid  \eta^2 \mid \psi \rangle
-\langle \psi \mid  \eta \mid \psi \rangle^2   \Bigr)
\end{align}
with
\begin{equation}\label{etaG}
\eta=\eta(\alpha(l))=\frac{dU(\alpha(l))}{dl}U^{\dagger}(\alpha(l))
=G_{i}(\alpha(l))\dot{\alpha}^{i},
\end{equation}
provided that the following relation has been used:
\begin{equation}
\frac{d}{dl} \left( \frac{\partial \eta}{\partial \dot{\alpha}^{i}}\right)
=\frac{\partial \eta}{\partial \alpha^{i}}
 +\left[ \eta,\ \frac{\partial \eta}{\partial \dot{\alpha}^{i}}\right].
\end{equation}
	
Up to this stage, we have purely discussed the variational 
problem for the functional in Eq. (\ref{functional}), 
which is naturally assumed to be analytic, 
and have not used the flow equation yet.
	
We are now going to prove that under a certain condition 
the generalized flow equation (\ref{generalized flow equation}) 
makes Eq. (\ref{X}) vanish and defines a geodesic curve 
associated with the ``initial state'' 
$\mid \psi \rangle=\mid u_{n} \rangle$, which is an 
eigenstate of $H_{\text{d}}(l)$. 
(This is the state of relevance, because we are considering 
the flow to an exact stationary state.) That is, we are 
going to show that, under the condition found later, 
the above $X_{i}$ vanishes for  $\eta=\eta^{(a)}$ 
in Eq. (\ref{ath eta}).
	
Clearly, $\langle u_{n} \mid \eta^{(a)} \mid u_{n} \rangle$ 
and its derivatives with respect to $l$ 
and $\dot{\alpha}^{i}$ vanish. Furthermore, it can be shown 
by using Eqs. (\ref{diagonal})-(\ref{ath eta}) that
$\langle u_{n} \mid $ 
$\left[\eta^{(a)^2},\ \partial\eta^{(a)}/\partial \dot{\alpha}^{i}\right]$
$\mid u_{n} \rangle$  
also vanishes. Therefore, $X_{i}$ for $\mid \psi \rangle=\mid u_{n} \rangle$ 
and $\eta=\eta^{(a)}$ is reduced to
\begin{equation} \label{reduced X}
X_{i}=2\langle u_{n} \mid \eta^{(a)^2} \mid u_{n} \rangle
 \langle u_{n} \mid 
\left\{ \frac{d \eta^{(a)}}{dl},\  
\frac{\partial \eta^{(a)}}{\partial \dot{\alpha}^{i}} \right\}
\mid u_{n} \rangle 
-\langle u_{n} \mid \frac{d \eta^{(a)^2}}{dl} \mid u_{n} \rangle
\langle u_{n} \mid \frac{\partial \eta^{(a)^2}}{\partial \dot{\alpha}^{i}}
\mid u_{n} \rangle.
\end{equation}
This equation can be rewritten by using 
Eqs. (\ref{diagonal})-(\ref{ath eta}) again as follows:
\begin{align} \label{rewritten X}
X_{i}=\ &4
\left( 
\Bigl|D^{(a)}_{n+i_{a}}\Bigr|^2 +\Bigl|D^{(a)}_{n}\Bigr|^2 
\sum_{m=1}^d \delta_{n,m+i_{a}}
\right)
\text{Re}
\left[ 
\frac{d D^{(a)}_{n+i_{a}}}{dl}
 \frac{\partial D^{(a)^*}_{n+i_{a}}}{\partial \dot{\alpha}^{i}}
+\frac{d D^{(a)}_{n}}{dl}
 \frac{\partial D^{(a)^*}_{n}}{\partial \dot{\alpha}^{i}}
   \sum_{m=1}^d \delta_{n,m+i_{a}}
\right] \nonumber \\
&-\left( 
\frac{d \Bigl|D^{(a)}_{n+i_{a}}\Bigr|^2}{dl}
 +\frac{d \Bigl|D^{(a)}_{n}\Bigr|^2}{dl}
\sum_{m=1}^d \delta_{n,m+i_{a}}     
\right)
\left( 
\frac{\partial \Bigl|D^{(a)}_{n+i_{a}}\Bigr|^2 }{\partial \dot{\alpha}^{i}}
 +\frac{\partial \Bigl|D^{(a)}_{n}\Bigr|^2}{\partial \dot{\alpha}^{i}}
  \sum_{m=1}^d \delta_{n,m+i_{a}}
\right),
\end{align}
where 
$D^{(a)}_{n+i_{a}} \equiv 
\Bigl( \epsilon_{n+i_{a}}(l)-\epsilon_{n}(l) \Bigr)
C^{(a)}_{n+i_{a}}(l)$. 
The $l$-derivatives appearing in this expression should be 
calculated from the generalized flow equation 
in Eq. (\ref{generalized flow equation}). 
It is also noted that $\alpha^{i}$ and $\dot{\alpha}^{i}$ are
not independent in Eqs. (\ref{reduced X}) and 
(\ref{rewritten X}) any more, 
since Eq. (\ref{ath eta}) has already been used.
	
Now, we present our discovery. 
If the labels $a$ and $a^{\prime}$ satisfying
\begin{equation} \label{condition}
i_{a^{\prime}} \neq 2i_{a}, 3i_{a} \quad (1\leq a^{\prime} \leq A)
\end{equation}
can be taken for $\eta^{(a)}(l)$, 
then sandwiching the flow equation 
in Eq. (\ref{generalized flow equation}) 
by three pairs,  
$\langle u_{n} \mid$ and $\mid u_{n-i_{a}} \rangle$,
$\langle u_{n+i_{a}} \mid$ and $\mid u_{n} \rangle$,
$\langle u_{n+i_{a}} \mid$ and $\mid u_{n-i_{a}}\rangle$,
we have
\begin{equation}
\langle u_{n} \mid 
\left[ \eta^{(a)}(l),\ \sum_{a^{\prime}=1}^A H^{(a^{\prime})}_{\text{o-d}}(l)  
\right]
\mid u_{n-i_{a}} \rangle=0,
\end{equation}
\begin{equation}
\langle u_{n+i_{a}} \mid 
\left[ \eta^{(a)}(l),\ \sum_{a^{\prime}=1}^A H^{(a^{\prime})}_{\text{o-d}}(l)  
\right]
\mid u_{n} \rangle=0,
\end{equation}
\begin{equation}
\langle u_{n+i_{a}} \mid 
\left[ \eta^{(a)}(l),\ \sum_{a^{\prime}=1,\ a^{\prime}\neq a}^A 
H^{(a^{\prime})}_{\text{o-d}}(l)  
\right]
\mid u_{n-i_{a}} \rangle=0,
\end{equation}
respectively. And, correspondingly, we obtain the 
following sandwiched flow equations:
\begin{equation} \label{Cn}
\frac{d C^{(a)}_{n}(l)}{dl}
=-\Bigl( \epsilon_{n}(l)-\epsilon_{n-i_{a}}(l)\Bigr)^2 C^{(a)}_{n}(l),
\end{equation}
\begin{equation} \label{Cni}
\frac{d C^{(a)}_{n+i_{a}}(l)}{dl}
=-\Bigl( \epsilon_{n+i_{a}}(l)-\epsilon_{n}(l)\Bigr)^2 C^{(a)}_{n+i_{a}}(l),
\end{equation}
\begin{equation} \label{epsilon}
\Bigl(
 \epsilon_{n+i_{a}}(l)+\epsilon_{n-i_{a}}(l)-2\epsilon_{n}(l)
\Bigr)
C^{(a)}_{n}(l) C^{(a)}_{n+i_{a}}(l)=0.
\end{equation}
It is mentioned that, from 
Eqs. (\ref{Cn}) and (\ref{Cni}), 
the phases of $C^{(a)}_{n}(l)$ and $C^{(a)}_{n+i_{a}}(l)$ 
are found to be independent of $l$.

Case-A: \quad If $C^{(a)}_{n}(l)=C^{(a)}_{n+i_{a}}(l)=0$, 
		       then $X_{i}$ in Eq. (\ref{rewritten X}) obviously 
		       vanishes. 

Case-B: \quad If $C^{(a)}_{n}(l)=0$ and $C^{(a)}_{n+i_{a}}(l)\neq 0$
              (or, $C^{(a)}_{n}(l)\neq 0$
               and $C^{(a)}_{n+i_{a}}(l)= 0$), then $X_{i}$ 
               vanishes, 

\hspace{1.6cm}due to the fact that the phases 
of $D^{(a)}_{n}(l)$ and $D^{(a)}_{n+i_{a}}(l)$ 
are independent of $l$. 
		
Case-C: \quad If both $C^{(a)}_{n}(l)$ and $C^{(a)}_{n+i_{a}}(l)$ 
		       are nonzero, then Eq. (\ref{epsilon}) 
		       yields $\epsilon_{n+i_{a}}(l)+\epsilon_{n-i_{a}}(l)-2\epsilon_{n}(l)=0$. \\
Combining the relation in Case-C with 
Eqs. (\ref{Cn}) and (\ref{Cni}), we obtain a crucial result 
that $\Bigl|C^{(a)}_{n+i_{a}}(l)\Bigr|$ 
\textit{is proportional to} $\Bigl|C^{(a)}_{n}(l)\Bigr|$. 
And, this makes $X_{i}$ vanish again.
	
Therefore, we conclude that the flow equation with the 
generator $\eta^{(a)}(l)$ with the label $a$ satisfying 
the condition in Eq. (\ref{condition}) gives rise to 
the geodesic flow of $\mid u_{n} \rangle$, independently of 
a specific initial condition. 
This is the main result of the present work.
	
We emphasize that Eq. (\ref{condition}) does not necessarily 
lead to the geodesic nature of Wegner's flow generated 
by $\eta^{W} \equiv \sum_{a=1}^A \ \eta^{(a)}$: 
it guarantees the geodesic nature of the flow generated 
by $\eta^{(a)}$. However, according to our experience, 
there exist physically important examples, in which $A=1$, 
that is, the present method becomes reduced to Wegner's. 
So, in such a case, Eq. (\ref{condition}) works for 
establishing the geodesic nature of Wegner's flow. 
In the next section, some of such examples are discussed.

\section{Physical Examples}
\label{sec:Examples}

\ \ \ \ \ \ In what follows, we illustrate our result by analyzing some physical examples.

The first example we consider is the generalized harmonic 
oscillator. The Hamiltonian reads
\begin{equation}
H=\omega a^{\dagger}a+\lambda a^{\dagger^2}+\lambda^* a^2
  +\mu a^{\dagger}+\mu^* a+\nu.
\end{equation}
Here, $a^{\dagger}$ and $a$ are the creation and 
annihilation operators obeying the algebra: $[a,a^{\dagger}]=1$, 
$[a^{\dagger},a^{\dagger}]=[a,a]=0$. $\omega$ $(>0)$ 
and $\nu$ are real constants, whereas $\lambda$ and $\mu$ 
are complex. In particular, the condition, $\omega>2|\lambda|$, 
is to be satisfied. Here and hereafter, $\hslash$ is set equal 
to unity. We analyze the following two cases: 
(i) $\mu \neq 0$, $\lambda=0$ and 
(ii) $\mu= 0$, $\lambda \neq 0$. 
[In the case when both $\mu$ and $\lambda$ are nonzero, 
one can first eliminate the terms linear 
in $a^{\dagger}$ and $a$ by the unitary transformation 
with the displacement operator, 
$\text{exp}(\alpha a^{\dagger}-\alpha^{*}a)$, 
where $\alpha$ is a complex $c$-number variable 
satisfying $\omega \alpha + 2\lambda \alpha^{*}-\mu=0$. 
Therefore, this case is reduced to (ii).] 
In case (i), the Hamiltonian is transformed by 
the displacement operator
\begin{equation}
U(l)=\text{exp}\Bigl[z(l)a^{\dagger}-z^{*}(l)a \Bigr]
\end{equation}
with the complex $z(l)$ satisfying $z(0)=0$. 
For the normalized ground state, $\mid 0 \rangle$, 
satisfying $a\mid 0 \rangle=0$, as the ``initial state'' 
at $l=0$, the flow of the state is along the coherent state. 
The corresponding Fubini-Study metric is Euclidean \cite{Abe}:
\begin{equation}
ds^2=\frac{1}{2}(dx^2+dp^2)
\end{equation}
with the parametrization, $z=(x+ip)/\sqrt{2}$. 
Then, the condition in Eq. (\ref{condition}) is satisfied, 
and the flow is, in fact, geodesic: a straight line in 
the space with a global coordinate $(x,p)$. 
In case (ii), the squeezing operator
\begin{equation}
U(l)=\text{exp} \left[
\frac{1}{2}\left( \xi(l)a^{\dagger^2}-\xi^{*}(l)a^2 \right) 
\right]
\end{equation}
is to be considered, where the complex coefficient $\xi(l)$ is 
parametrized as  $\xi(l)=r(l)e^{-2i\phi(l)}$
 $( 0\leq r(l),0 \leq \phi(l)<2 \pi )$. 
Accordingly, $(\alpha^1,\alpha^2) \equiv (r,\phi)$. 
The condition  $U(0)=I$ (with the identity operator $I$) 
leads to $r(0)=0$. The initial state is taken to be 
the number state, $\mid n \rangle =(n!)^{-1/2}a^{\dagger^n} \mid 0 \rangle$. 
The corresponding metric is \cite{Abe}:
\begin{equation}
ds^2=\frac{1}{2}(n^2+n+1)
\Bigl[ dr^2+\left( \text{sinh}^2 2r\right)d\phi^2 \Bigr],
\end{equation}
which shows that the manifold is the Lobachevsky space. 
The transformed Hamiltonian is written as follows:
\begin{equation}
H(l)=\omega(l) a^{\dagger}a+\lambda(l) a^{\dagger^2}+\lambda^*(l) a^2+\nu(l).
\end{equation}
The coefficients appearing here depend not only on their 
original values at $l=0$ but also on $\xi(l)$. 
In this case,
\begin{equation}
H_{\text{o-d}}(l)=\lambda(l) a^{\dagger^2}+\lambda^*(l) a^2
\end{equation}
is the one and only off-diagonal part. Therefore, 
the generator in Eq. (\ref{ath eta}) is identical to 
Wegner's choice
\begin{equation}
\eta^{W}(l)=2\omega(l) \left[\lambda(l) a^{\dagger^2}-\lambda^*(l) a^2 \right],
\end{equation}
and the condition in Eq. (\ref{condition}) is automatically 
fulfilled. The flow equation for $\lambda(l)$ is given by
\begin{equation}
\frac{d \lambda(l)}{d l}=-4\omega^2(l)\lambda(l),
\end{equation}
showing that the phase of $\lambda(l)$ does not depend on $l$. 
Then, one can explicitly find that Eq. (\ref{reduced X}) 
indeed vanishes. Comparing $\bigl[dU(l)/dl\bigr]U^{\dagger}(l)$ 
with $\eta^{W}(l)$, we obtain $\phi(l)=\text{const}$, 
which in fact turns out to make both $\delta S/\delta r(l)$ 
and $\delta S/\delta \phi(l)$ vanish, where $S$ is 
the arc-length functional defined in terms of the above metric. 
We also mention that the spectrum 
of $H_{\text{d}}(l)=\omega(l)a^{\dagger}a+\nu(l)$ is equally 
spaced, and accordingly Case-C in the preceding section is 
realized. Thus, Wegner's flow is geodesic. 
	
The second example is a spin-$s$, 
$\vec{S}=(S_{x},S_{y},S_{z})$, in a constant external 
magnetic field $\vec{B}$. 
The Hamiltonian reads
\begin{equation}
H=\vec{S} \cdot \vec{B}
\end{equation}
in an appropriate unit. The unitary operator to be considered 
is
\begin{equation}
U(l)=\text{exp} \bigl[ \sigma(l)S_{+}-\sigma^{*}S_{-} \bigr],
\end{equation}
where $S_{\pm} \equiv S_{x} \pm i S_{y}$ and 
the complex coefficient $\sigma(l)$ is parametrized 
as $\sigma(l)=[ \theta(l)/2]e^{-i \varphi(l)}$ 
$( 0 \leq \theta(l)< \pi, 0 \leq \varphi(l)<2\pi)$ 
with $\theta(0)=0$. The basic commutation relations satisfied 
by the spin operators are 
as follows: $\bigl[S_{z},S_{\pm}\bigr]=\pm S_{\pm}$, 
$\bigl[S_{+},S_{-}\bigr]=2S_{z}$. The local coordinate is given 
by $(\alpha^1,\alpha^2) \equiv (\theta,\varphi)$. 
The initial state is taken to be 
$\mid m \rangle_{s}$
$(m=-s,-s+1,\ldots,0,\ldots,s-1,s)$, which satisfies 
$S_{z} \mid m \rangle_{s}=m \mid m \rangle_{s}$. 
The metric is found to be given by \cite{Abe}:
\begin{equation}
ds^2=\frac{1}{2}(s^2+s-m^2)
 \Bigl[ d\theta^2+\left( \text{sin}^2\theta  \right)d\varphi^2 \Bigr],
\end{equation}
which is of a sphere. The transformed Hamiltonian is written 
as
\begin{equation}
H(l)=\beta_{z}(l) S_{z}+\beta(l)S_{+}+\beta^{*}(l)S_{-}.
\end{equation}
$\beta_{z}(l)$ is real, whereas $\beta(l)$ is complex. 
They depend not only on $\vec{B}$ but also on 
$\sigma(l)$. The one and only off-diagonal part is
\begin{equation}
H_{\text{o-d}}(l)=\beta(l)S_{+}+\beta^{*}(l)S_{-},
\end{equation}
and so Wegner's choice
\begin{equation}
\eta^{W}(l)=\beta_{z}(l) 
\bigl[ \beta(l)S_{+}-\beta^{*}(l)S_{-} \bigr]
\end{equation}
is employed. Therefore, clearly the condition 
in Eq. (\ref{condition}) is fulfilled. 
The flow equation for $\beta(l)$ is
\begin{equation}
\frac{d \beta(l)}{d l}=-\beta^{2}_{z}(l)\beta(l),
\end{equation}
from which the phase of $\beta(l)$ is seen to be independent 
of $l$, and accordingly Eq. (\ref{reduced X}) vanishes. 
Comparison of the above generator, $\eta^{W}(l)$, 
with $\bigl[d U(l)/dl\bigr]U^{\dagger}(l)$ 
yields $\varphi(l)=\text{const}$, which leads to the fact 
that the variations of the arc-length functional 
($S$, calculated by using the above metric) with respect 
to $\theta(l)$ and $\varphi(l)$ vanish. 
Also, the spectrum of $H_{\text{d}}(l)=\beta_{z}(l)S_{z}$ 
with $s \geq 1$ is equally spaced, and so Case-C in the 
preceding section is realized. 
Thus, Wegner's flow is geodesic (i.e., the great circle).
	
The third example is the Jaynes-Cummings model 
\cite{Jaynes}, which describes a two-level atom interacting 
with a single-mode radiation field. 
The Hamiltonian is given by
\begin{equation}
H=\frac{1}{2}\omega_{0}\sigma_{3}+\omega a^{\dagger}a
  +\kappa(\sigma_{+}a+\sigma_{-}a^{\dagger}).
\end{equation}
Here, $\omega_{0}$, $\omega$, and $\kappa$ are the transition 
frequency of the atom, the frequency of the radiation, 
and the coupling constant, respectively. $a^{\dagger}$ and $a$ 
are the creation and annihilation operators of the radiation 
field satisfying the same algebra as in the first example. 
$\sigma$'s are the operators of the atom, which are given 
in terms of the orthonormal basis, 
the ground state $\mid g \rangle$ and 
the excited state $\mid e \rangle$, 
as follows: $\sigma_{+}=\mid e \rangle \langle g \mid$, 
$\sigma_{-}=\mid g \rangle \langle e \mid$, 
$\sigma_{3}=\mid e \rangle \langle e \mid-\mid g \rangle \langle g \mid$. 
This model is known to be exactly solvable. 
The associated unitary operator is
\begin{equation}
U(l)=\sum_{n=0}^{\infty}
\Bigl[ \alpha_{n}(l) \sigma_{+}\sigma_{-} \mid n \rangle \langle n \mid 
+\beta_{n-1}(l) \sigma_{-}\sigma_{+} \mid n \rangle \langle n \mid 
+\gamma_{n}(l)\sigma_{+} \mid n \rangle \langle n+1 \mid
+\delta_{n}(l)\sigma_{-} \mid n+1 \rangle \langle n \mid \Bigl],
\end{equation}
where $\mid n \rangle=(n!)^{-1/2}a^{\dagger^n} \mid 0 \rangle$ 
is the $n$-photon state 
(for the details of this unitary operator, see \cite{Itto}). 
The unitarity of this operator leads to the conditions: 
$\bigl|\alpha_{n}\bigr|^2+\bigl|\gamma_{n}\bigr|^2=1$, 
$\bigl|\alpha_{n}\bigr|=\bigl|\beta_{n}\bigr|$, 
$\bigl|\gamma_{n}\bigr|=\bigl|\delta_{n}\bigr|$, 
and $\alpha_{n} \delta^{*}_{n}+\beta^{*}_{n} \gamma_{n}=0$. 
From the polar forms, 
$\alpha_{n}=\bigl|\alpha_{n}\bigr|\text{exp}\left(i \theta_{\alpha_{n}}\right)$ 
and so on, follows the condition on 
the phases: $\theta_{\alpha_{n}}+\theta_{\beta_{n}}-\theta_{\gamma_{n}}$
$-\theta_{\delta_{n}}$ $=(2m+1)\pi$
$(m=0,\pm 1,\pm 2, \ldots)$. Here, we choose 
$\mid e \rangle \mid n \rangle$
 as the initial state. $\bigl|\alpha_{n}\bigr|$, 
$\theta_{\alpha_{n}}$, $\theta_{\beta_{n}}$, and 
$\theta_{\gamma_{n}}$ form as a set of independent 
coordinate variables. The metric is
\begin{equation}
ds^2=\frac{d \bigl|\alpha_{n}\bigr|^2}{1-\bigl|\alpha_{n}\bigr|^2}
  +\bigl|\alpha_{n}\bigr|^2 \left(1-\bigl|\alpha_{n}\bigr|^2\right) 
\Bigl(d \theta_{\alpha_{n}}-d \theta_{\gamma_{n}}\Bigr)^2,
\end{equation}
which does not seem to be the one of a familiar space. 
The $\theta_{\beta_{n}}$-dependence disappears due to 
the above choice of the initial state. 
The transformed Hamiltonian is
\begin{equation}
H(l)=\sum_{n=0}^{\infty} 
\Bigl[ A_{n}(l) \sigma_{+}\sigma_{-} \mid n \rangle \langle n \mid 
+B_{n-1}(l) \sigma_{-}\sigma_{+} \mid n \rangle \langle n \mid 
+C_{n}(l)\sigma_{+} \mid n \rangle \langle n+1 \mid
+C^{*}_{n}(l)\sigma_{-} \mid n+1 \rangle \langle n \mid \Bigl],
\end{equation}
and so
\begin{equation}
H_{\text{o-d}}(l)=\sum_{n=0}^{\infty}
\Bigl[ C_{n}(l)\sigma_{+} \mid n \rangle \langle n+1 \mid
+C^{*}_{n}(l)\sigma_{-} \mid n+1 \rangle \langle n \mid \Bigl],
\end{equation}
where the coefficients, $A_{n}(l)$'s and $B_{n-1}(l)$'s 
are real, and $C_{n}(l)$'s turn out to be also real 
(as can be seen from the flow equations). 
They are expressed in terms of the physical coefficients 
contained in the original Hamiltonian as well as the 
coefficients appearing in $U(l)$ 
(for details, see \cite{Itto}). 
Let us consider Wegner's choice:
\begin{equation}
\eta^{W}(l)=\sum_{n=0}^{\infty} \Bigl[A_{n}(l)-B_{n}(l) \Bigr]
C_{n}(l)\bigl( \sigma_{+} \mid n \rangle \langle n+1 \mid
-\sigma_{-} \mid n+1 \rangle \langle n \mid \bigr).
\end{equation}
Then, we find that Eq. (\ref{reduced X}) vanishes. 
The flow equations give rise to
\begin{equation}
\frac{d \theta_{\alpha_{n}}(l)}{d l}=0,
\end{equation}
\begin{equation}
\frac{d \theta_{\gamma_{n}}(l)}{d l}=0,
\end{equation}
which explicitly make the variations of the arc-length 
functional with respect to $\bigl|\alpha_{n}(l)\bigr|$, 
$\theta_{\alpha_{n}}(l)$, and $\theta_{\gamma_{n}}(l)$ 
all vanish. 
In addition, comparing 
$\eta(l)=\bigl[H_{\text{d}}(l),\ H_{\text{o-d}}(l)\bigr]$ with 
$\eta^{W}(l)$ above, 
we see that Case-B in the preceding section is realized. 
Thus, again Wegner's flow is geodesic.

\section{Concluding Remarks}
\label{sec:Concluding Remarks}

\ \ \ \ \ To summarize, generalizing Wegner's method of flow equations, 
we have found a condition, under which the corresponding flow 
of a quantum state becomes geodesic in a quantum evolution 
submanifold, independently of specific initial 
conditions at $l=0$. 
We have illustrated this by employing the generalized 
harmonic oscillator, the spin in an external magnetic field, 
and the Jaynes-Cummings model.

The present result implies that the method of flow equations 
is not just a mathematical tool for diagonalizing a 
Hamiltonian but provides the optimal strategy in quantum 
state engineering for realizing a stationary state from 
a given initial state in each of a certain class of systems. 
In fact, a formal solution, 
$U(l)= P_{l} \ \text{exp} \int_{0}^{l}dl^{\prime}\eta^{W}(l^{\prime})$ 
with $P_{l}$ being Dyson's ``$l$-ordering'' symbol, 
can be divided into the product of many unitary operators, 
each of which defines a small translation along the geodesic 
flow and may represent each quantum operation performed 
experimentally. In the present work, we have discussed 
the metric structure and associated geodesic nature of 
the flow of a state in a quantum evolution submanifold. 
Other basic quantities such as curvature are not explicitly 
treated here. However, curvature becomes relevant, 
for example, when geodesic deviation is considered. 
This kind of considerations may cast further light on 
geometry of the flow equations.
	
All the examples discussed in Section 4 are exactly solvable, 
and Wegner's method perfectly works, generating the geodesic 
flows. This observation naturally leads to an anticipation 
that Wegner's flow may always be geodesic if a system is 
exactly solvable. Still there does not exist an 
affirmative/negative proof of this point, and further 
investigations are needed.



\end{document}